\documentclass[a4paper]{article}

\usepackage{url}
\usepackage{INTERSPEECH2020}
\usepackage{flushend}
\usepackage{cite}
\usepackage{epstopdf}

\title{Shouted Speech Compensation for Speaker Verification Robust to Vocal Effort Conditions}

\name{Santi Prieto$^1$, Alfonso Ortega$^2$, Iv\'an L\'opez-Espejo$^3$, Eduardo Lleida$^2$}
\address{
  $^1$VeriDas $\vert$ das-Nano, Navarre, Spain\\
  $^2$ViVoLab, Arag\'{o}n Institute for Engineering Research (I3A)\\
University of Zaragoza, Spain\\
    $^3$Department of Electronic Systems, Aalborg University, Denmark}
\email{sprieto@das-nano.com, \{ortega,lleida\}@unizar.es, ivl@es.aau.dk}

\graphicspath{{Images/}}

\begin{document}

\maketitle
\thispagestyle{empty}
\pagestyle{empty}

\begin{abstract}
The performance of speaker verification systems degrades when vocal effort conditions between enrollment and test (e.g., shouted \emph{vs.} normal speech) are different. This is a potential situation in non-cooperative speaker verification tasks. In this paper, we present a study on different methods for linear compensation of embeddings making use of Gaussian mixture models to cluster shouted and normal speech domains. These compensation techniques are borrowed from the area of robustness for automatic speech recognition and, in this work, we apply them to compensate the mismatch between shouted and normal conditions in speaker verification. Before compensation, shouted condition is automatically detected by means of logistic regression. The process is computationally light and it is performed in the back-end of an x-vector system. Experimental results show that applying the proposed approach in the presence of vocal effort mismatch yields up to 13.8\% equal error rate relative improvement with respect to a system that applies neither shouted speech detection nor compensation.
\end{abstract}

\section{Introduction}

Nowadays, the development of speaker verification systems has become very popular in real applications and it will continue growing over the next years. Many researchers are focused on the normal speech situation, an area where these systems begin to work very well. Nevertheless, very few studies address situations where there is an extreme mismatch in terms of vocal effort. Vocal effort has been studied in \cite{Zhang2007,Shriberg2008}, showing differences in five different modes: whisper, soft, normal, loud and shout. Each mode alters speech production in a way that introduces noticeable differences that affect the performance of the speaker verification systems. For this reason, we consider that it is necessary to study each mode in order to be able to automatically detect it, and perform an appropriate pre-processing to compensate the changes in the features and the negative impact on performance of non-neutral modes.

This work is intended to study the effect of shouted speech in speaker verification systems that are trained with normal speech, detecting and compensating this mismatch in order to improve the speaker verification performance. State-of-the-art speaker verification systems obtain very good performance in the normal speech scenario (e.g., \cite{Jung18}), when utterances are in the same collaborative conditions (neutral, normal, calm environment, etc.). Some works \cite{Shriberg2008,Pohjalainen2012} study the effect of vocal effort conditions on the performance of speaker verification systems and demonstrate that the accuracy is negatively affected in the presence of vocal effort mismatch.

High vocal effort mode is used by speakers to produce a louder acoustic signal in order to increase the signal-to-noise ratio usually in noisy environments, distant communication or in emergency situations. There are some works that address the issue of how to detect high vocal effort \cite{Pohjalainen2012, Pohjalainen2013, Chao2018}. In \cite{Pohjalainen2013}, spectral characteristics of shouted speech are studied in order to demonstrate that many acoustic properties of the voice change in high vocal effort situations. In this paper, we focus on detecting shouted speech as a previous step to alleviate the performance degradation of speaker verification systems in those situations. Our detection method is based on a logistic regression model trained on embeddings directly obtained from shouted and normal utterances by using a time-delay neural network (TDNN) as described in \cite{Snyder2018}.

Finally, we propose a method based on \cite{Buera2007} to reduce the error when embeddings extracted from shouted utterances are used in speaker verification systems trained with normal speech data. In \cite{Buera2007}, a feature compensation technique for speech recognition in noisy domains is presented. Mel-frequency cepstral coefficients (MFCCs) are used to train different Gaussian mixture models (GMMs) associated to different noisy conditions and, then, bias compensation terms are estimated depending on the noisy environment. Finally, acoustic features are compensated with these terms to improve speech recognition performance in noisy conditions. In this work, we adapt this technique to train GMMs using embeddings extracted by employing a TDNN-based model instead of MFCCs and compensating shouted embeddings in order to be able to mitigate the negative effect of shouted speech on the performance of speaker verification systems. We demonstrate that applying a linear compensation approach like this in the presence of vocal effort mismatch yields a relative improvement of up to 13.8\% in terms of equal error rate (EER) in comparison with a system that applies neither shouted speech detection nor compensation.

The remainder of this paper is organized as follows: a brief comparison between shouted and normal speech is given in Section \ref{sec:normal_vs_shouted}. Shouted speech detection is described in Section \ref{sec:detection}. Section \ref{sec:compensation} deals with shouted speech compensation. The experimental setup and results are presented in Sections \ref{sec:setup} and \ref{sec:results}, respectively. Finally, Section \ref{sec:conclusions} concludes this work.

\section{Shouted \emph{vs.} Normal Speech}
\label{sec:normal_vs_shouted}

Many works have analyzed the acoustic differences between shouted and normal speech \cite{Zhang2007,Shriberg2008,Lopez2017}. For instance, the authors of \cite{Lopez2017} demonstrated that the increase of vocal effort changes many acoustical properties of speech. In the spectral domain, it makes the fundamental frequency and the first formant to increase, as well as flattening of the spectral tilt. Hence, short-term spectral features such as MFCCs are thus directly affected by the increased vocal effort, which in turn affects the speaker recognition performance. To mitigate this effect, a spectral matching between shouted and normal speech on a perceptual scale was proposed in \cite{Lopez2017}.

In this work, we want to study how these differences between shouted and normal speech can affect the speaker verification performance. State-of-the-art speaker verification is based on a TDNN trained with MFCCs to obtain speaker embeddings \cite{Snyder2018}. Differences between shouted and normal speech also affect both the intra- and inter-speaker variability at the TDNN output, which can be visualized in the embedding domain. For that, i.e., to transform the embeddings extracted from the TDNN and see how shouted and normal speech conditions are represented in a two-dimensional space, we use t-SNE \cite{vanDerMaaten2008}.

\begin{figure}[t]
\centering
     \includegraphics[width=0.72\columnwidth]{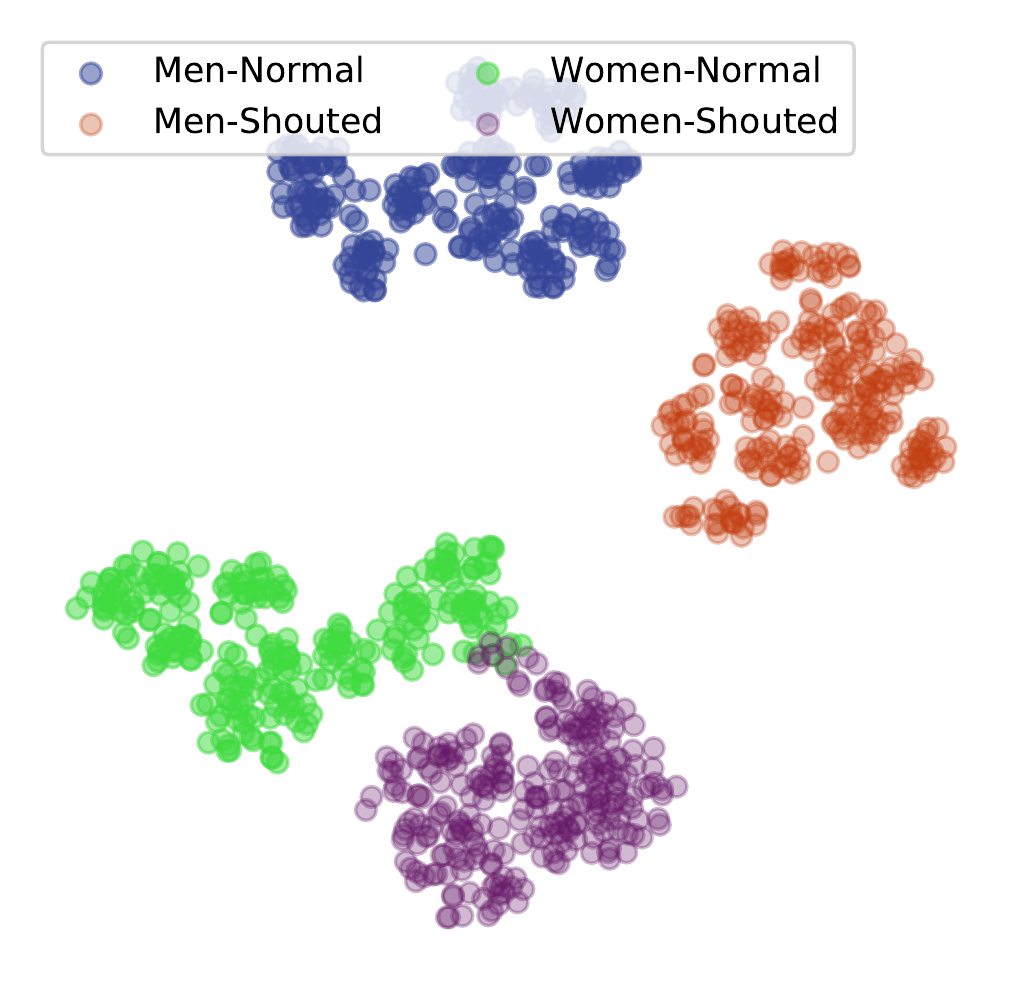}
     \caption{A comparison between speaker embeddings (projected onto a two-dimensional space using t-SNE) extracted from shouted and normal speech.}
     \label{fig:NormalShoutedTSNE}
\end{figure}

In Figure \ref{fig:NormalShoutedTSNE}, a two-dimensional speaker embedding representation from 11 males and 11 females is shown. Each speaker is characterized by 24 shouted and 24 normal points in the two-dimensional space. We can observe four clusters that represent different embedding characteristics. On the one hand, there are gender representation clusters for embeddings, and, on the other hand, there are shouted and normal speech differences in the embedding characteristics that are visualized in distinct clusters. This affects the speaker verification task due to the intra-speaker variability introduced by the two vocal effort domains, one of shouted utterances and the other one of normal speech. If two utterances from the same speaker but with different conditions are compared, the system will not be able to verify them correctly as same speaker because the embeddings are affected by vocal effort mismatch and will be rejected.

\section{Shouted Speech Detection}
\label{sec:detection}

To avoid introducing unnecessary distortion to normal speech embeddings, it is crucial to develop an accurate shouted speech detector before applying the proposed compensation techniques. To this end, we assume this task as a two-class classification problem by training a logistic regression model with embeddings obtained from shouted and normal speech utterances. Logistic regression is chosen due to both low complexity and very good performance. Let $\mathbf{z}=\left(z_1,...,z_D\right)^\top$ be a $D$-dimensional embedding, $H_0$ and $H_1$ indicate the hypotheses that $\mathbf{z}$ is a shouted and a normal speech embedding, respectively. Thus, the probability that $\mathbf{z}$ comes from shouted speech, $P(H_0|\mathbf{z})=1-P(H_1|\mathbf{z})$, is estimated in this work as
\begin{equation}
 P(H_0|\mathbf{z})=\frac{1}{1+\exp{\left\{-\left(\beta_0+\beta_1z_1+\cdots+\beta_Dz_D\right)\right\}}},
\end{equation}
where, as aforementioned, the parameters of the model, $\left\{\beta_j;\;j=0,...,D\right\}$, are calculated from a set of training embeddings obtained from shouted and normal speech (see Subsection \ref{ssec:test}). At test time, an embedding $\mathbf{z}$ is classified as coming from a shouted speech utterance if $P(H_0|\mathbf{z})>0.5$.

The usefulness of this rather simple, yet effective method is shown in the result section.
\section{Shouted Speech Compensation}
\label{sec:compensation}

In this section, we describe the technique used to compensate the shouted speech embeddings. This technique is simple and has only a few parameters to better fit the data scarcity. We propose here the use of \emph{Multi-Environment Model-based LInear Normalization} (MEMLIN) \cite{Buera2007}, a method borrowed from robust speech recognition. Given the normal speech embedding $\mathbf{x}$ and the shouted one $\mathbf{y}$, a normal speech embedding estimate, $\hat{\mathbf{x}}$, can be obtained by minimum mean square error estimation as
\begin{equation}
\hat{\mathbf{x}}=E[\mathbf{x}|\mathbf{y}] = \int{\mathbf{x}\cdot p(\mathbf{x}|\mathbf{y})d\mathbf{x}},
\label{eq:MMSE}
\end{equation}
where $E[\cdot]$ is the expectation operator and $p(\mathbf{x}|\mathbf{y})$ is the conditional probability density function of $\mathbf{x}$ given $\mathbf{y}$. In order to evaluate the expression in Eq. (\ref{eq:MMSE}) for MEMLIN, the following assumptions are made.

First, normal speech embeddings are modelled by using a GMM:
\begin{equation}
    p(\mathbf{x})=\sum_{s_x} p(\mathbf{x}|s_x)P(s_x),
\label{eq:GMMx}
\end{equation}
with 
\begin{equation}
p(\mathbf{x}|s_x)=\mathcal{N}({\mathbf{x}|\boldsymbol\mu_{s_x},\boldsymbol\Sigma_{s_x}}),
\label{eq:Gaussx}
\end{equation}
where $s_x$ denotes each Gaussian of the normal speech model, and $\boldsymbol\mu_{s_x}$, $\boldsymbol\Sigma_{s_x}$ and $P(s_x)$ are the mean, covariance matrix (which is diagonal in this work as we assume statistical independence among embedding components) and weight associated to Gaussian $s_x$. In addition, $p(\mathbf{x}|s_x)$ is the likelihood of the normal speech embedding given the Gaussian $s_x$.

Secondly, shouted speech embeddings are similarly modelled as
\begin{equation}
p(\mathbf{y})=\sum_{s_y}p(\mathbf{y}|s_y)P(s_y),
\label{eq:GMMy}
\end{equation} 
with
\begin{equation}
p(\mathbf{y}|s_y) = \mathcal{N}(\mathbf{y}|\boldsymbol\mu_{s_y},\boldsymbol\Sigma_{s_y}).
\label{eq:Gaussy}
\end{equation}

Finally, the third assumption is considering that the normal embedding, $\mathbf{x}$, can be obtained from the shouted embedding, $\mathbf{y}$, by making use of the above models:
\begin{equation}
\mathbf{x}=\mathbf{f}(\mathbf{y},s_x,s_y).
\label{eq:normal_to_shouted_regression}
\end{equation}

With all of these assumptions, Eq. (\ref{eq:MMSE}) can be expressed by using the Bayes' rule and the proposed models for both domains as
\begin{equation}
\begin{array}{ccc}
    \hat{\mathbf{x}} & = & \displaystyle\int \sum_{s_y} \sum_{s_x} \mathbf{x} \cdot p(\mathbf{x},s_x,s_y|\mathbf{y}) \cdot p(s_x,s_y|\mathbf{y})d\mathbf{x} \vspace{0.1cm} \\
     & \approx & \mathbf{y}-\displaystyle\sum_{s_y} \sum_{s_x} \mathbf{r}_{s_x s_y} \cdot p(s_y|\mathbf{y}) \cdot p(s_x|\mathbf{y},s_y),
\end{array}
\end{equation} 
where $\mathbf{r}_{s_x s_y}$ is a bias term (see below). Given the shouted speech embedding $\mathbf{y}$, to obtain an estimate $\hat{\mathbf{x}}$ of the normal speech embedding it is necessary to compute the probability of the shouted speech Gaussian $s_y$ given $\mathbf{y}$, $p(s_y|\mathbf{y})$, and the probability of the normal speech Gaussian $s_x$ given the shouted embedding $\mathbf{y}$ and the shouted speech Gaussian $s_y$, $p(s_x|\mathbf{y},s_y)$.

The bias terms, $\mathbf{r}_{s_x s_y}$, are obtained in a training stage using a set of paired embeddings (see Subsection \ref{ssec:test}) from both domains, $\left\{\mathbf{x}_i^{T\!r},\mathbf{y}_i^{T\!r}\right\}$, following
\begin{equation}
\mathbf{r}_{s_x s_y} = \frac {\sum_i {p\left(s_y,\mathbf{y}_i^{T\!r}\right)p\left(s_x,\mathbf{x}_i^{T\!r}\right)\left(\mathbf{y}_i^{T\!r}-\mathbf{x}_i^{T\!r}\right)}} {\sum_i {p\left(s_y,\mathbf{y}_i^{T\!r}\right)p\left(s_x,\mathbf{x}_i^{T\!r}\right)}}.
\label{eq:bias_term}
\end{equation} 

In order to compare the performance of the proposed method with some other well-known compensation techniques for robustness in automatic speech recognition, we also implemented two techniques such as \emph{Multivariate Gaussian-based Cepstral Normalization} (RATZ) \cite{Moreno96} and \emph{Stereo-based Piecewise LInear Compensation for Environments} (SPLICE) \cite{Droppo01}.

In RATZ, the normal speech embedding is modelled using a GMM in the normal speech domain according to Eqs. (\ref{eq:GMMx}) and (\ref{eq:Gaussx}), and the estimation of $\mathbf{x}$ follows
\begin{equation}
\hat{\mathbf{x}}= \mathbf{f}_{R\!A\!T\!Z}(\mathbf{y},s_x) \approx \mathbf{y} - \sum_{s_x} {\mathbf{r}_{s_{x}} \cdot p(s_x|\mathbf{y})},
\label{eq:RatzApprox}
\end{equation}
where $\mathbf{r}_{s_{x}}$ is a bias term that only depends on the normal speech Gaussian and is obtained in a previous training phase from the set of paired embeddings $\left\{\mathbf{x}_i^{T\!r},\mathbf{y}_i^{T\!r}\right\}$ according to
\begin{equation}
\mathbf{r}_{s_{x}} = \frac{\sum_i{p\left(s_x\left|\mathbf{x}_i^{T\!r}\right)\right.\left(\mathbf{y}_i^{T\!r}-\mathbf{x}_i^{T\!r}\right)}}{\sum_i{p\left(s_x\left|\mathbf{x}_i^{T\!r}\right)\right.}}.
\label{eq:Ratz_Bias}
\end{equation} 

On the other hand, in SPLICE, the shouted speech domain is modelled with a GMM according to Eqs. (\ref{eq:GMMy}) and (\ref{eq:Gaussy}), and the estimate of $\mathbf{x}$ is obtained following
\begin{equation}
\hat{\mathbf{x}}= \mathbf{f}_{S\!P\!L\!I\!C\!E}(\mathbf{y},s_y) \approx \mathbf{y} - \sum_{s_y} {\mathbf{r}_{s_{y}} \cdot p(s_y|\mathbf{y})},
\label{eq:splice_approx}
\end{equation} 
where $\mathbf{r}_{s_{y}}$ is a bias term obtained in a training stage again using the set of paired embeddings from both domains as follows:
\begin{equation}
\mathbf{r}_{s_{y}} = \frac{\sum_i{p\left(s_y\left|\mathbf{y}_i^{T\!r}\right)\right.\left(\mathbf{y}_i^{T\!r}-\mathbf{x}_i^{T\!r}\right)}}{\sum_i{p\left(s_y\left|\mathbf{y}_i^{T\!r}\right)\right.}}.
\label{eq:Splice_Bias}
\end{equation} 

\begin{figure}[t]
\centering
     \includegraphics[width = 0.72 \columnwidth]{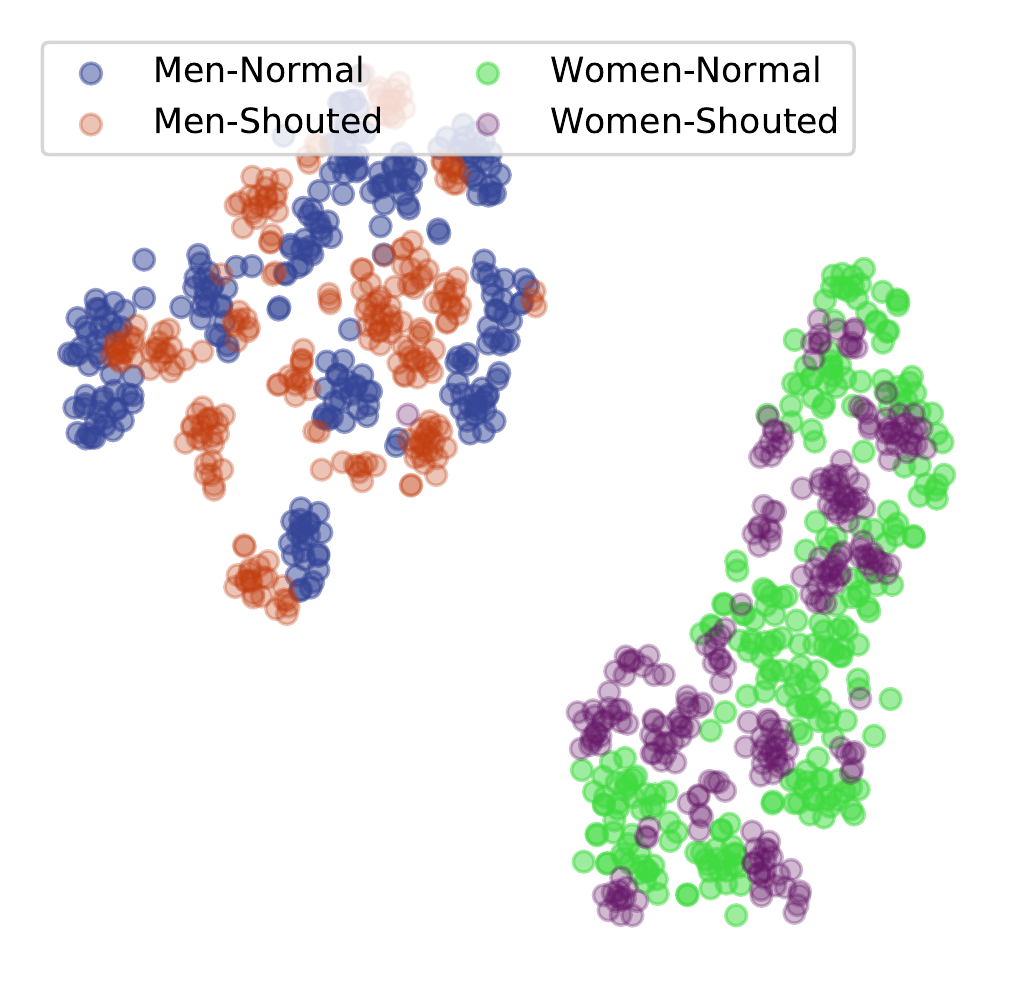}
     \caption{A comparison between shouted and normal speech embeddings (projected onto a two-dimensional space using t-SNE) when MEMLIN is employed for shouted embedding compensation.}
     \label{fig:NormalShoutedTSNE_memlin}
\end{figure}

\section{Experimental Setup}
\label{sec:setup}

\subsection{Speaker Verification System}
\label{ssec:svs}

The speaker verification system is implemented according to the x-vector-based Kaldi \cite{Povey2011} recipe using augmented versions of the VoxCeleb1 \cite{VC1} and VoxCeleb2 \cite{VC2} corpora\footnote{\url{https://github.com/kaldi-asr/kaldi/tree/master/egs/voxceleb}}. The models generated from this recipe are freely available on the Internet\footnote{\url{https://kaldi-asr.org/models/m7}}. The EER (which is the primary evaluation metric in this paper) obtained using this baseline system for VoxCeleb is 3.1\%.

This speaker verification system consists of a TDNN-based front-end for 512-dimensional speaker embedding (x-vector) computation (i.e., $D=512$) plus a probabilistic linear discriminant analysis (PLDA) back-end for verification. The TDNN is fed with 30-dimensional MFCC features extracted from speech signals that are framed using a 25 ms analysis window with a 10 ms shift. Voice activity detection is employed to discard non-speech frames. Then, prior PLDA scoring, x-vectors are centered, reduced in terms of dimensionality by means of linear discriminant analysis and length-normalized.

\begin{figure*}
    \centering
    \includegraphics[width=0.67\columnwidth]{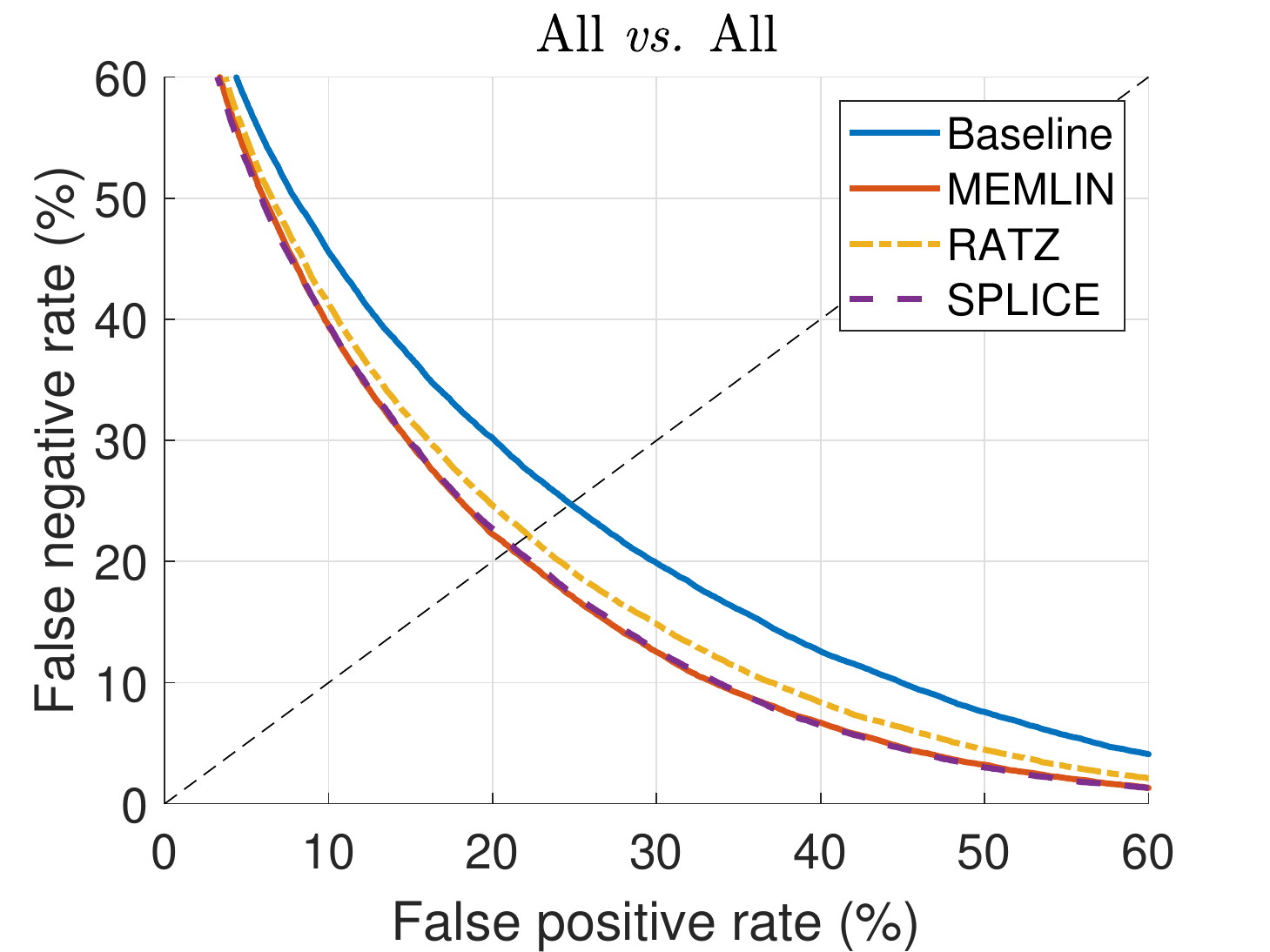} \includegraphics[width=0.67\columnwidth]{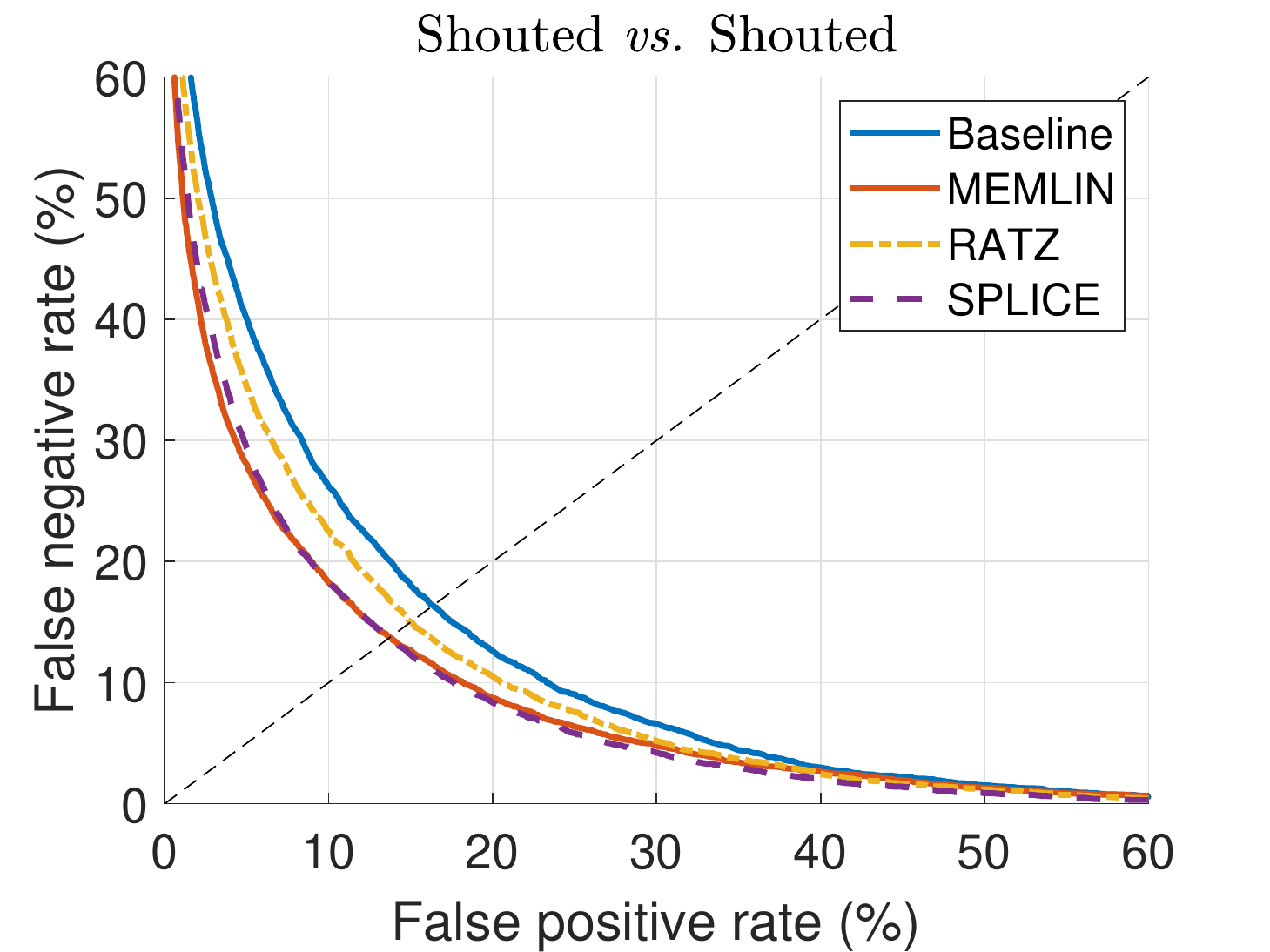} \includegraphics[width=0.67\columnwidth]{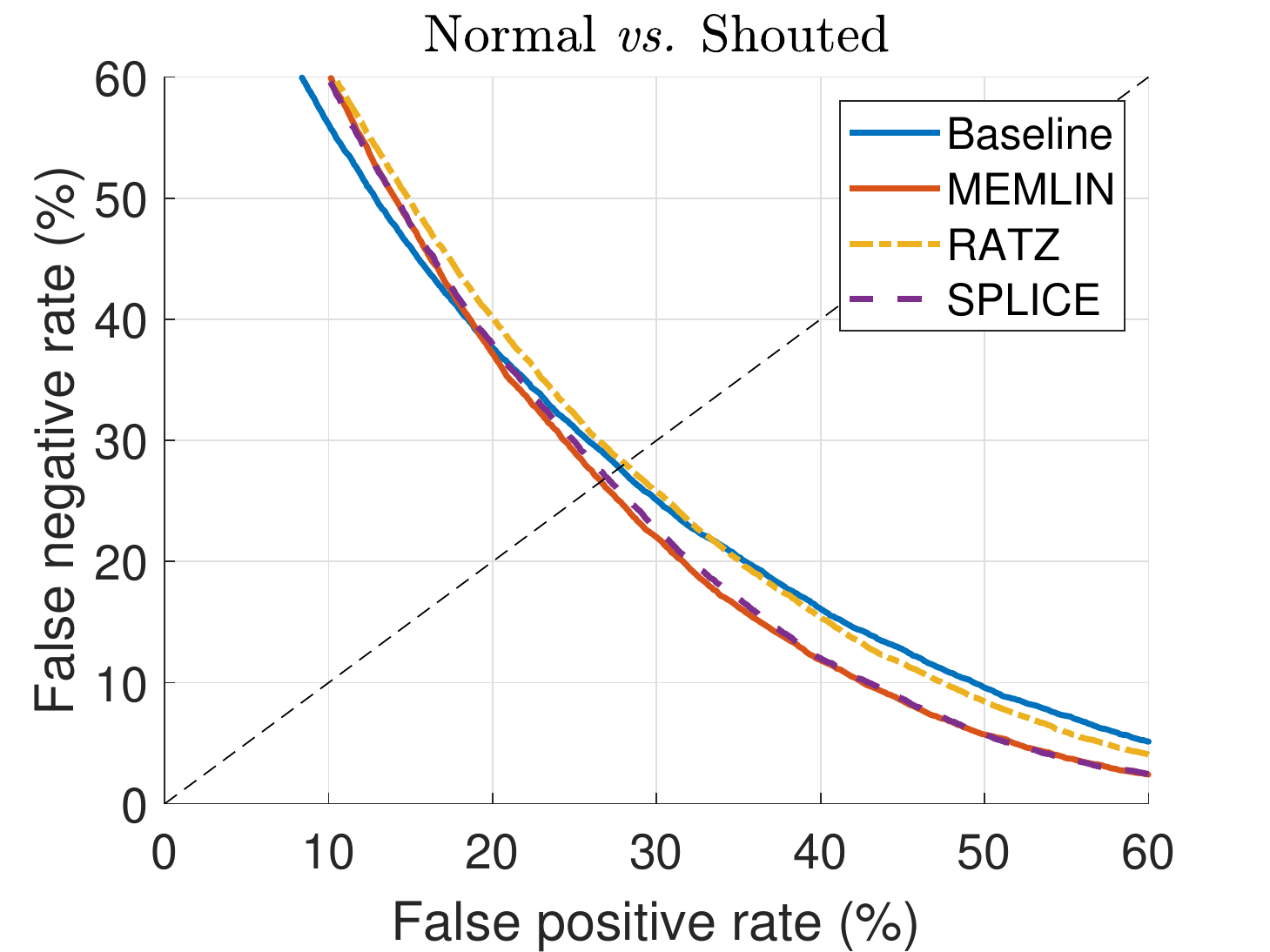}
    \caption{Detection error trade-off curves for the techniques tested in this work when using the proposed shouted speech detection. Different plots refer to different conditions. From left to right: All \emph{vs.} All, Shouted \emph{vs.} Shouted and Normal \emph{vs.} Shouted conditions.}
    \label{fig:det_curves}
\end{figure*}

\subsection{Test Database}
\label{ssec:test}

The speech corpus used to perform the experiments is the one presented in \cite{Hanilci2013}. It consists of 11 male and 11 female speakers. Each of them recorded 24 sentences speaking normally and the same 24 sentences shouting. The sentences were recorded in an anechoic chamber using a high-quality microphone. Channel effects and environment variations were completely excluded. The sentences were spoken in Finnish, half in imperative and half in indicative mode. The average duration of each utterance is 3 seconds.

Due to the scarcity of shouted speech, both shouted speech detection and compensation experiments are carried out using leave-one-speaker-out cross-validation to maximize the number of trials. All the utterances in the corpus are processed to extract x-vectors according to the process outlined in Subsection \ref{ssec:svs} and further detailed in \cite{Snyder2018}. Four different conditions are considered for experimental evaluation:

\begin{itemize}
  \item \textbf{All \emph{vs.} All (A-A)}: All the shouted and normal speech utterances are compared each other, which yields 557,040 verification trials.
  \item \textbf{Normal \emph{vs.} Normal (N-N)}: Normal speech utterances are compared each other, which yields 139,128 verification trials.
  \item \textbf{Shouted \emph{vs.} Shouted (S-S)}: Shouted speech utterances are compared each other, which yields 139,128 verification trials.
  \item \textbf{Normal \emph{vs.} Shouted (N-S)}: Normal speech utterances are compared against shouted speech utterances, which yields 278,784 verification trials.
\end{itemize}


\section{Results}
\label{sec:results}

In this section, the use of MEMLIN-, RATZ- and SPLICE-based shouted speech compensation, also considering the proposed shouted speech detection, is compared with a baseline speaker verification system that applies neither shouted speech detection nor shouted speech compensation. Notice that the shouted speech compensation techniques employ 8-component GMMs.

\begin{table}
\centering
\begin{tabular}{ | c || c | c | c | c | }
\hline
\textbf{Condition} & \textbf{Baseline} & \textbf{MEMLIN} & \textbf{RATZ} & \textbf{SPLICE}\\
\hline
\hline
     A-A  & 24.76 & \textbf{21.09} & 22.16 & 21.24\\
     N-N  & 12.93 & 12.93 & 12.93 & 12.93\\
     S-S  & 16.37 & 13.73 & 15.00 & \textbf{13.60}\\
     N-S  & 27.73 & \textbf{26.60} & 28.06 & 26.95\\
\hline
\end{tabular}
\caption{Speaker verification results in terms of EER, in percentages, when using oracle shouted speech detection.}
\label{table:oracle}
\end{table}

First, Table \ref{table:oracle} shows speaker verification results in terms of EER, in percentages, when oracle shouted speech detection is used. As we can see, for Baseline there is a relative worsening of around 114\% between Normal \emph{vs.} Normal and Normal \emph{vs.} Shouted conditions that justifies the need for vocal effort mismatch compensation. To a greater or lesser extent, such a mismatch is reduced by MEMLIN, RATZ and SPLICE. More in particular, the best results are those obtained by MEMLIN and SPLICE in contrast to RATZ, which suggests that modelling the shouted embedding space is important to achieve better compensation performance. Furthermore, the utility of MEMLIN for shouted embedding compensation can be visually inspected in Figure \ref{fig:NormalShoutedTSNE_memlin}.

\begin{table}
\centering
\begin{tabular}{ | c|| c | c | c | c | }
\hline
\textbf{Condition} & \textbf{Baseline} & \textbf{MEMLIN} & \textbf{RATZ} & \textbf{SPLICE}\\
\hline
\hline
     A-A  & 24.76 & 21.62 & 22.26 & \textbf{21.34}\\
     N-N  & 12.93 & 12.93 & 13.14 & 13.20\\
     S-S  & 16.37 & 14.03 & 15.15 & \textbf{13.76}\\
     N-S  & 27.73 & \textbf{26.43} & 28.17 & 27.00\\
\hline
\end{tabular}
\caption{Speaker verification results in terms of EER, in percentages, when using the proposed shouted speech detection.}
\label{table:shout_detection}
\end{table}

Similarly to Table \ref{table:oracle}, Table \ref{table:shout_detection} shows speaker verification results when using the shouted speech detection proposed in Section \ref{sec:detection}. These results are supported by the detection error trade-off curves of Figure \ref{fig:det_curves}. Considering, indeed, leave-one-speaker-out cross-validation, our shouted speech detector obtains 98.11\% accuracy, where only 1.17\% and 2.65\% of shouted and normal utterances, respectively, are misclassified. In these circumstances, it is not surprising the high similarity between the results reported in Tables \ref{table:oracle} and \ref{table:shout_detection}. It is important to remark that, in the more interesting from a practical perspective All \emph{vs.} All scenario, SPLICE achieves a 13.8\% relative improvement with respect to Baseline (in accordance with Table \ref{table:shout_detection}).

\begin{table}
\centering
\begin{tabular}{ | c|| c | c | c | c | }
\hline
\textbf{Condition} & \textbf{Baseline} & \textbf{MEMLIN} & \textbf{RATZ} & \textbf{SPLICE}\\
\hline
\hline
     A-A  & 24.76 & 21.92 & 22.84 & \textbf{21.80}\\
     N-N  & 12.93 & 12.93 & 12.93 & 13.20\\
     S-S  & 16.37 & 14.75 & 18.70 & \textbf{14.59}\\
     N-S  & 27.73 & 27.61 & 27.49 & \textbf{27.32}\\
\hline
\end{tabular}
\caption{Gender-dependent speaker verification results in terms of EER, in percentages and averaged across genders, when using oracle shouted speech detection.}
\label{table:Gender-Dependent}
\end{table}

From Figure \ref{fig:NormalShoutedTSNE}, one may think that applying gender-dependent shouted embedding compensation can bring about an improvement with respect to employing a gender-independent approach. For this reason, we evaluated gender-dependent versions of MEMLIN-, RATZ- and SPLICE-based shouted speech compensation, the results (averaged across genders) of which are shown in Table \ref{table:Gender-Dependent}. As can be seen, the equivalent gender-independent shouted embedding compensation of Table \ref{table:oracle} is superior to the gender-dependent approach.
 

\section{Conclusions}
\label{sec:conclusions}

In this work, we have shown the need for vocal effort mismatch compensation in the context of speaker verification. Moreover, we have also shown the potential of several linear compensation techniques intended to mitigate the mismatch between speaker embeddings extracted from shouted and normal speech utterances. These techniques have worked on top of a very effective shouted speech detector based on logistic regression.

As there is certainly room for improvement, future work will be concerned with studying other mismatch compensation approaches possibly involving unsupervised learning or transfer learning. Towards this goal, we will require the acquisition of larger corpora comprising high vocal effort speech data.

\section{Acknowledgements}
Authors would like to thank Dr. Tomi Kinnunen for providing the database we have performed this study with. This work has been partially supported by the Spanish Ministry of Economy and Competitiveness and the European Social Fund through the project TIN2017-85854-C4-1-R, Government of Arag\'{o}n (Reference Group T36\_17R) and co-financed with Feder 2014-2020 ``Building Europe from Arag\'{o}n''.

\bibliographystyle{IEEEtran}

\bibliography{mybib}

\begin{thebibliography}{10}
\providecommand{\url}[1]{#1}
\csname url@samestyle\endcsname
\providecommand{\newblock}{\relax}
\providecommand{\bibinfo}[2]{#2}
\providecommand{\BIBentrySTDinterwordspacing}{\spaceskip=0pt\relax}
\providecommand{\BIBentryALTinterwordstretchfactor}{4}
\providecommand{\BIBentryALTinterwordspacing}{\spaceskip=\fontdimen2\font plus
\BIBentryALTinterwordstretchfactor\fontdimen3\font minus
  \fontdimen4\font\relax}
\providecommand{\BIBforeignlanguage}[2]{{%
\expandafter\ifx\csname l@#1\endcsname\relax
\typeout{** WARNING: IEEEtran.bst: No hyphenation pattern has been}%
\typeout{** loaded for the language `#1'. Using the pattern for}%
\typeout{** the default language instead.}%
\else
\language=\csname l@#1\endcsname
\fi
#2}}
\providecommand{\BIBdecl}{\relax}
\BIBdecl

\bibitem{Zhang2007}
C.~Zhang and J.~Hansen, ``Analysis and classification of speech mode: Whispered
  through shouted,'' in \emph{Proc. of 8th Annual Conference of the
  International Speech Communication Association (INTERSPEECH)}, 2007, pp.
  2289--2292.

\bibitem{Shriberg2008}
E.~Shriberg, M.~Graciarena, H.~Bratt, A.~Kathol, S.~Kajarekar, H.~Jameel,
  C.~Richey, and F.~Goodman, ``Effects of vocal effort and speaking style on
  text-independent speaker verification,'' in \emph{Proc. of 9th Annual
  Conference of the International Speech Communication Association
  (INTERSPEECH)}, 2008, pp. 609--612.

\bibitem{Jung18}
J.-W. Jung, H.-S. Heo, I.-H. Yang, H.-J. Shim, and H.-J. Yu, ``A complete
  end-to-end speaker verification system using deep neural networks: From raw
  signals to verification result,'' in \emph{Proc. of 43rd International
  Conference on Acoustics, Speech and Signal Processing (ICASSP)}, 2018, pp.
  5349--5353.

\bibitem{Pohjalainen2012}
J.~Pohjalainen, T.~Raitio, H.~Pulakka, and P.~Alku, ``Automatic detection of
  high vocal effort in telephone speech,'' in \emph{Proc. of 13th Annual
  Conference of the International Speech Communication Association
  (INTERSPEECH)}, 2012.

\bibitem{Pohjalainen2013}
J.~Pohjalainen, T.~Raitio, S.~Yrttiaho, and P.~Alku, ``Detection of shouted
  speech in noise: Human and machine,'' \emph{The Journal of the Acoustical
  Society of America}, vol. 133, pp. 2377--2389, 2013.

\bibitem{Chao2018}
H.~Chao, L.~Dong, and Y.~Liu, ``Two-stage vocal effort detection based on
  spectral information entropy for robust speech recognition,'' \emph{Journal
  of Information Hiding and Multimedia Signal Processing}, vol.~9, pp.
  1496--1505, 2018.

\bibitem{Snyder2018}
D.~Snyder, D.~Garcia-Romero, G.~Sell, D.~Povey, and S.~Khudanpur, ``X-vectors:
  Robust {DNN} embeddings for speaker recognition,'' in \emph{Proc. of 43rd
  International Conference on Acoustics, Speech and Signal Processing
  (ICASSP)}, 2018, pp. 5329--5333.

\bibitem{Buera2007}
L.~Buera, E.~Lleida, A.~Miguel, A.~Ortega, and O.~Saz, ``Cepstral vector
  normalization based on stereo data for robust speech recognition,''
  \emph{IEEE Transactions on Audio, Speech, and Language Processing}, vol.~15,
  pp. 1098--1113, 2007.

\bibitem{Lopez2017}
A.~Lopez, R.~Saeidi, L.~Juvela, and P.~Alku, ``Normal-to-shouted speech
  spectral mapping for speaker recognition under vocal effort mismatch,'' in
  \emph{Proc. of 42nd International Conference on Acoustics, Speech and Signal
  Processing (ICASSP)}, 2017, pp. 4940--4944.

\bibitem{vanDerMaaten2008}
\BIBentryALTinterwordspacing
L.~van~der Maaten and G.~Hinton, ``Visualizing data using {t-SNE},''
  \emph{Journal of Machine Learning Research}, vol.~9, pp. 2579--2605, 2008.
  [Online]. Available: \url{http://www.jmlr.org/papers/v9/vandermaaten08a.html}
\BIBentrySTDinterwordspacing

\bibitem{Moreno96}
P.~Moreno, ``Speech recognition in noisy environments,'' Ph.D. dissertation,
  ECE Department, Carnegie-Mellon University, 1996.

\bibitem{Droppo01}
J.~Droppo, L.~Deng, and A.~Acero, ``Evaluation of the {SPLICE} algorithm on the
  {Aurora2} database,'' in \emph{Proc. of 7th European Conference on Speech
  Communication and Technology (EUROSPEECH)}, 2001, pp. 217--220.

\bibitem{Povey2011}
D.~Povey, A.~Ghoshal, G.~Boulianne, L.~Burget, O.~Glembek, N.~Goel,
  M.~Hannemann, P.~Motlíček, Y.~Qian, P.~Schwarz, J.~Silovský, G.~Stemmer,
  and K.~Vesel, ``The {Kaldi} speech recognition toolkit,'' 2011.

\bibitem{VC1}
A.~Nagrani, J.~S. Chung, and A.~Zisserman, ``{VoxCeleb}: a large-scale speaker
  identification dataset,'' in \emph{Proc. of 18th Annual Conference of the
  International Speech Communication Association (INTERSPEECH)}, 2017, pp.
  2616--2620.

\bibitem{VC2}
J.~S. Chung, A.~Nagrani, and A.~Zisserman, ``{VoxCeleb2}: Deep speaker
  recognition,'' in \emph{Proc. of 19th Annual Conference of the International
  Speech Communication Association (INTERSPEECH)}, 2018, pp. 1086--1090.

\bibitem{Hanilci2013}
C.~{Hanilçi}, T.~{Kinnunen}, R.~{Saeidi}, J.~{Pohjalainen}, P.~{Alku}, and
  F.~{Ertaş}, ``Speaker identification from shouted speech: Analysis and
  compensation,'' in \emph{38th International Conference on Acoustics, Speech
  and Signal Processing (ICASSP)}, 2013, pp. 8027--8031.

\end{thebibliography}

\end{document}